\documentstyle[aps,multicol,epsfig]{revtex}

\begin{document}
\draft

\title{Evidence for saturation of channel transmission from conductance 
fluctuations in atomic-size point contacts}
\author{B. Ludoph$^{1}$, M.H. Devoret$^{2}$, D. Esteve$^{2}$, C. Urbina$^{2}$ 
and 
J.M van  Ruitenbeek$^{1}$ }
\address{$^{1}$ Kamerlingh Onnes Laboratorium, Leiden University, Postbus 9504, 
2300 RA Leiden, The Netherlands \newline
$^{2}$ Service de Physique de l'Etat Condense, CEA-Saclay, 91191 Gif-sur-Yvette, 
France}

\maketitle

\begin{abstract}
The conductance of atomic size contacts has a small, random, voltage dependent 
component analogous to conductance fluctuations observed in diffusive wires 
(UCF). A new effect is observed in gold contacts, consisting of a marked 
suppression of these fluctuations when the conductance of the contact is close 
to integer multiples of the conductance quantum. Using a model based on the 
Landauer-B\"uttiker formalism we interpret this effect as evidence that the 
conductance tends to be built up from fully transmitted (i.e., saturated) 
channels plus a single, which is partially transmitted.
\end{abstract}



Metallic contacts consisting of only a few atoms can be obtained using scanning 
tunneling microscopy (STM) or mechanically controllable break-junction 
\cite{MCB} techniques. The electrical conductance through such contacts is 
described in terms of electronic wave modes by the Landauer-B\"uttiker formalism 
\cite{landauer}. Each of the $N$ modes forms a channel for the conductance, with 
a transmission probability $T_{n}$ between 0 and 1. The total conductance is 
given by the sum over these channels $G = \sum_{n=1}^{N} T_{n} G_{0}$, with 
$G_{0} = 2e^{2}/h$ the quantum of conductance. By recording histograms of 
conductance values \cite{krans} for contacts of simple metals (Na, Au), a 
statistical preference was observed for conductances near integer values. This 
statistical preference was interpreted as an indication that transmitted modes 
in the most probable contacts are completely opened ($T_{n}=1$, i.e. saturation 
of channel transmission), in analogy with the conductance quantization observed 
in 2D electron gas devices \cite{2deg}. Here, we test this interpretation by performing a 
new type of measurement giving access to the second moment of the distribution 
of the $T_{n}$'s. We measure simultaneously, for a large number of gold atomic 
contacts obtained with the MCB technique, the conductance and its derivative 
with respect to the voltage.
\begin{figure}
\begin{center}
\leavevmode
\epsfig{figure=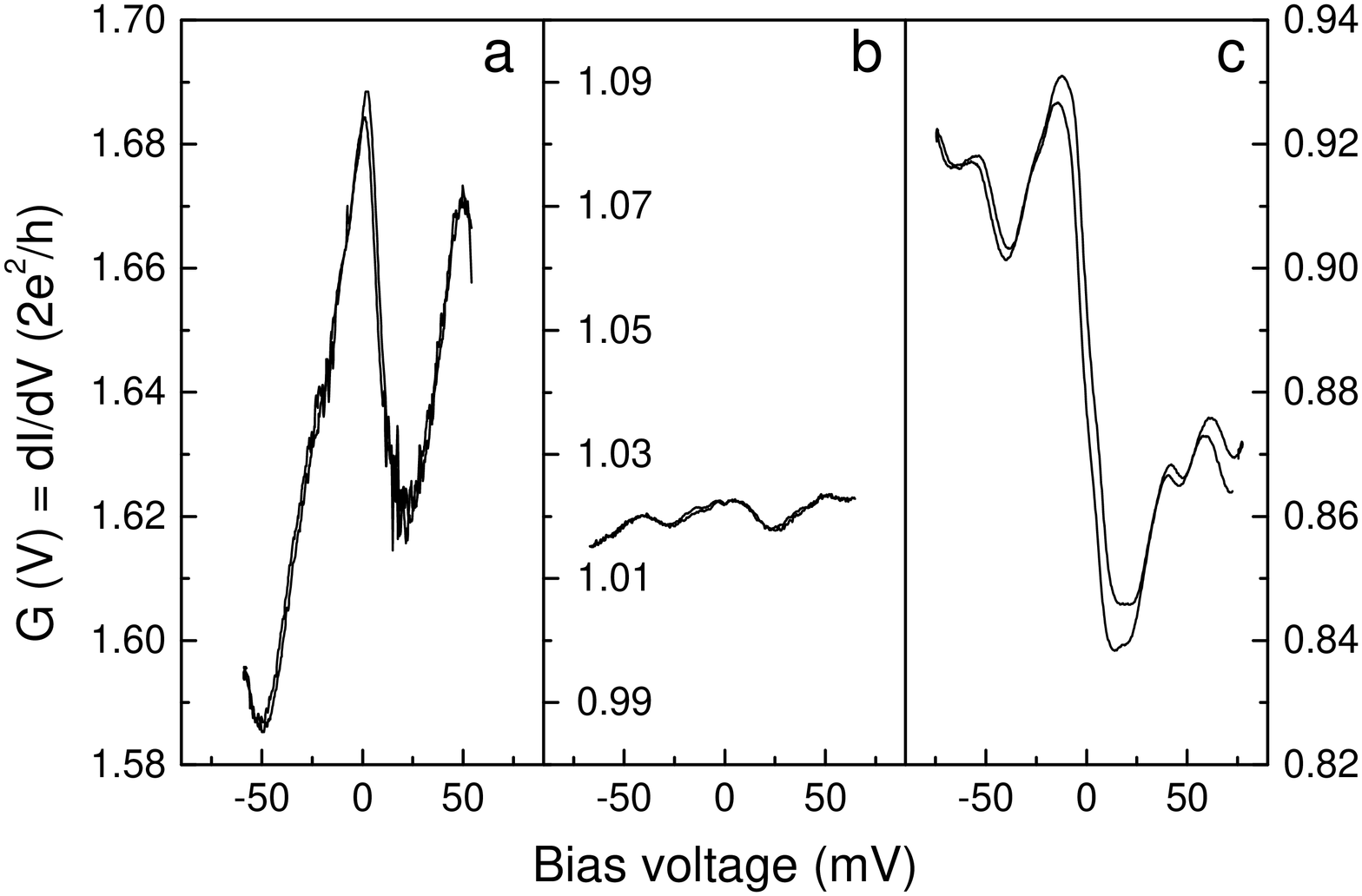,height=10cm,angle=270}
\end{center}
\caption{Differential conductance dI/dV as a function of bias voltage, measured 
with a modulation amplitude $<$0.35 mV, for three different contacts with  
$G\sim 1.65 G_{0}$ (a),  $\sim 1.02 G_{0}$ (b), and  $\sim 0.88 G_{0}$ (c). For all three curves the y-scale spans 0.12 $G_{0}$.}
\label{fig:fig1} 
\end{figure}
The atomic contacts are formed by breaking a gold wire at low temperatures, and 
then finely adjusting the size of the contact between the fresh fracture 
surfaces using a piezo-electric element \cite {MCB}. Fig.\,1 shows the differential 
conductance, $\partial I/\partial V$ measured as a function of bias voltage for 
three atomic size contacts with different conductance values, using a modulation 
voltage $eV \ll k_{B} \theta$ (with $\theta$ the temperature). For each contact, 
in order to illustrate reproducibility, both the curves for increasing and 
decreasing bias voltage are given. Measurements such as those of Fig.\,1 suggest 
that the fluctuation pattern changes randomly between contact configurations, 
and that the amplitude of the fluctuations is suppressed for conductance values 
near $G_0$. In order to establish such a relation it is necessary to 
statistically average over a large number of contacts. We do this by measuring 
the voltage dependence of the conductance ($\partial G/\partial V = \partial^{2} 
I/\partial V^{2}$) and the conductance itself ($G = \partial I/\partial V$) by 
applying a voltage modulation, and measuring the first and second harmonic of 
the voltage over a resistor in series with the contact. These are recorded 
continuously, while the contact is broken by increasing the voltage, $V_{P}$, 
over the piezo-electric element, producing curves as shown in Fig.\,2. We use a 
relatively large modulation amplitude of 20\,mV over the contact (at a frequency 
of 46\,kHz) in order to have sufficient sensitivity and speed of measurement, 
thus allowing averaging over many different contacts. The integration time of 
the lock-in amplifiers was 10\,ms and a reading was taken every 100\,ms. Between 
curves the contact was pushed together to a contact with conductance 
$>20\,G_{0}$, to ensure that a new contact geometry was measured each time. 
Frequently the contact required manual readjustment and, at these occasions, was 
pushed completely together. All measurements were performed on gold samples of 
6N purity, in vacuum at 4.2\,K. 
\begin{figure}
\begin{center}
\leavevmode
\epsfig{figure=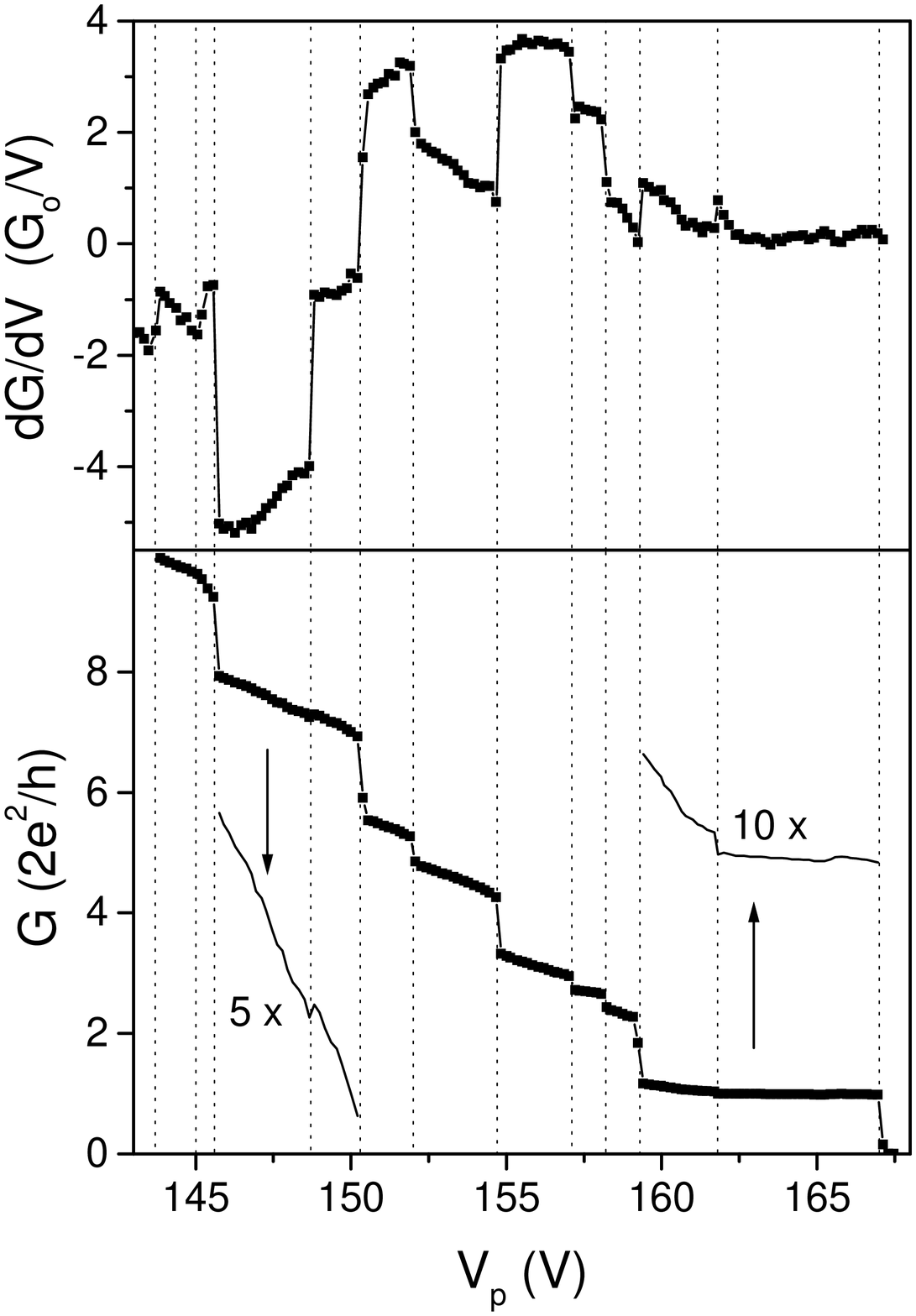,width=7cm,angle=0}
\end{center}
\caption{Typical example of the simultaneous measurement of voltage dependence 
of the conductance $\partial G/ \partial V$ and the conductance $G$, as a 
function of piezo voltage $V_{P}$. The graph includes vertical dotted lines that show that the steps in both quantities coincide. Two plateaus have been enlarged and offset to show the tiny steps in the conductance. The elongation of the contact is linear with $V_{P}$ and 10\,V corresponds to about 1\,nm.}
\label{fig:fig2} \end{figure}
The conductance in Fig.\,2 shows the typical behavior when breaking gold 
contacts \cite{goldsteps}, which consists of plateaus with steps of the order 
$G_{0}$ and a last plateau close to $G_{0}$ before entering the tunneling 
regime. The steps and plateaus in $G$ correspond with atomic rearrangements and 
elastic deformation respectively, as the contact is pulled apart and finally 
breaks \cite{agrait}. At each step in $G$ we find corresponding steps in 
$\partial G/ \partial V$. Even tiny steps in $G$, such as between 7 and 8 
$G_{0}$, can produce dramatic jumps in $\partial G/ \partial V$. 

$\partial G/ \partial V$ has a random sign and magnitude with a bell-shaped  
distribution. Fig.\,3a shows the standard deviation 
$\sigma_{GV}=\sqrt{\langle(\partial G/ \partial V)^{2}\rangle - \langle \partial 
G/ \partial V\rangle^{2}}$, as a function of $G$ together with a histogram of 
conductance values (Fig.\,3b) determined from 3500 individual curves similar to 
the one shown in Fig.\,2. The squares in Fig.\,3a represent the standard 
deviation for a fixed number of 2500 data points each, calculated after sorting 
the combined data of the 3500 curves as a function of conductance. Below 
0.8\,$G_{0}$ where very few points could be measured we use 300 data points 
(circles). Using fixed numbers of data points in calculating $\sigma_{GV}$ 
avoids influence of the histogram statistics (Fig.\,3b) on $\sigma_{GV}$. One 
clearly observes a very sharp minimum in Fig.\,3a at a conductance $G_{0}$ and 
less pronounced minima near 2, 3 and even 4\,$G_{0}$. This new observation forms 
the central result of this paper. Fig\,3a shows the combined results for three gold samples. The global features reproduce in all three cases, but some sample 
dependence is observed in the shape and height of the maxima. The histogram of 
conductance values (Fig.\,3b) is in accordance with previous measurements for 
gold at low temperature, e.g. \cite{sirvent}. 

The effect we observe has the same origin as that noted by Maslov {\it et al.} 
\cite{maslov} in numerical simulations on constrictions with defects. 
The principle can be understood by considering a contact with a single 
conducting mode having a finite transmission probability $T$, described by 
transmission and reflection coefficients $t$, $t^{\prime}$, $r$ and $r^{\prime}$ 
(coming in from left and right, respectively), with $|t^{\prime}|^{2}=|t|^2=T$ 
and  $|r^{\prime}|^{2}=|r|^2=1-T$. As illustrated in Fig.\,3 (inset), electron 
waves transmitted by the contact with amplitude $t$, and backscattered to the 
contact by diffusing paths with amplitude $a$, have a probability amplitude $r$ 
to be {\it reflected} at the contact. This wave interferes with the directly 
transmitted partial wave and modifies the total conductance. A similar 
contribution comes from the trajectories on the other side of the contact. These 
interference terms will be sensitive to changes in the phase accumulated along the 
trajectories, which is determined by the electron energy and the path length. We 
can change the energy by the applied voltage, giving rise to the fluctuations 
shown in Fig.\,1. Changes in path length of the order of the Fermi wavelength, 
which is the atomic scale, occur at the steps in the conductance in Fig.\,2, 
explaining the correlation with the steps in $\partial G/\partial V$. Each time 
when the contact is opened, and closed again to sufficiently large conductance 
values, random atomic reconfigurations take place, leading to a completely new 
set of scattering centers. Thus the result presented in Fig.\,3 can be 
interpreted as the ensemble average over defect configurations.

In the following paragraphs, we derive an analytical expression for $\sigma_{GV}$ to 
lowest order in $a$. In our model, the system is divided into a ballistic 
central constriction connected to diffusive conductors on each side (Fig.\,3, 
inset). The central part is described by a transfer matrix ${\bf t}$, with elements 
$t_{nm}$ giving the amplitude for the mode $n$ on the left to be transmitted 
into mode $m$ on the right of the constriction. After diagonalization only a few 
non-zero diagonal elements remain, corresponding to the number of conducting 
modes at the narrowest part of the conductor \cite{mads}. The $t_{nm}$ are 
energy dependent, but only on a very large energy scale so that we can ignore 
this in first approximation. For the total transmission of the combined system, 
in terms of the return amplitudes on the left and right hand side of the contact 
(${\bf a}_{l}(E)$ and ${\bf a}_{r}(E)$ respectively), we obtain the expression  
${\bf t}_{t}(E) = {\bf t}_{l} ({\bf t}^{\prime -1} - {\bf a}_{r}(E) {\bf t}^{\dagger  -1} {\bf a}_{l}(E)- {\bf a}_{r}(E) 
{\bf r}^{\prime} {\bf t}^{\prime -1}  - {\bf t}^{\prime -1} {\bf r} {\bf a}_{l}(E))^{-1} {\bf t}_{r}  $, with ${\bf r}$, 
${\bf r}'$ and ${\bf t}$, ${\bf t}'$ (now in the general multimode case) the matrices of 
reflection and transmission coefficients of the constriction. ${\bf t}_{l}$ and 
${\bf t}_{r}$ are the transmission matrices through the left and right diffusive 
regions, respectively. For $k_B \theta \ll eV$ the non-linear conductance can be 
expressed as $G=\partial I/\partial V$, 

\[ I = \frac{2e}{h} \int^{eV}_{0} Tr [{\bf t}_{t}(E){\bf t}^{\dagger}_{t}(E)] 
\text{d}E. \]

\noindent The fluctuations in the conductance are described by $\delta G = G - 
\langle G\rangle$ (where $\langle G\rangle$ is the conductance averaged over 
impurity configurations) of which we will consider the voltage dependence 
$\partial \delta G/ \partial V$. When we take into account that scattering 
processes in the left and right banks are uncorrelated, product terms of ${\bf a}_{l}$ 
and ${\bf a}_{r}$ disappear when we average. For the purpose of calculating the small 
fluctuating part of the conductance we can assume ${\bf t}_{l}{\bf t}_{l}^{\dagger}={\bf t}_{r}{\bf t}_{r}^{\dagger} \simeq 
\bf{1}$, although their deviation from unity will affect $\langle G\rangle$, 
which we will address briefly below. Considering first a contact with only a 
single transmitted mode, we obtain an expression for the voltage dependence of 
the conductance squared, averaged over impurity configurations:

\begin{eqnarray} 
\sigma_{GV}^{2} =  \Bigl< \Bigl( \frac{\partial \delta G}{\partial V}\Bigr)^{2} 
\Bigr> = G_{0}^{2} T^{2} (1-T) \nonumber \\
\times 2 \Bigl< \text{Re} \bigl(\frac{\partial a_{l}(eV)}{\partial V}
\frac{\partial a_{l}^{*}(eV)}{\partial V}  
+ \frac{\partial a_{r}(eV)}{\partial V}
\frac{\partial a_{r}^{*}(eV)}{\partial V} \bigr) \Bigr>. \label{flucts}
\end{eqnarray}

\noindent Products of the form $\langle a(E_{1})a^{*}(E_{2})\rangle$ can be 
expressed as $\int_{0}^{\infty} P_{cl}(\tau) e^{-i(E_{1}-E_{2})\tau/\hbar} 
\text{d}\tau $ in terms of the classical probability, $P_{cl}(\tau)= v_{F}/((1-
cos(\gamma))2 \sqrt{3 \pi} k_{F}^{2} (D\tau)^{3/2}) $ to return to the contact 
after a diffusion time $\tau$.
We assume the diffusion is into a cone of opening angle $\gamma $ (Fig.\,3, 
inset), $D = v_{F}l_{e}/3$ is the diffusion constant and $l_{e} = v_{F}\tau_{e}$ 
with $\tau_{e}$ the elastic scattering time \cite{daniel}. The differentiation 
of  $a(eV)$ in Eq.\,(\ref{flucts}) only affects the phase factors (to very good 
approximation), and produces a factor $(e\tau/\hbar)^2$ under the integral over 
the diffusion time, $\tau$. Further, taking into account that the finite 
modulation amplitude $V$ is the limiting energy scale ($k_{B}\theta, 
\hbar/\tau_{\phi} \ll eV$ with $\tau_{\phi}$ the inelastic scattering time), we 
obtain 

\begin{eqnarray} 
\sigma_{GV}^{2} = \Bigl( \frac {2.71\ e\  G_{0}}{\hbar k_{F}v_{F}
\sqrt{1-\cos\gamma}}\Bigr)^{2} 
\Bigl(\frac {\hbar/\tau_{e}}{eV}\Bigr)^{3/2}T^{2}(1-T).\label{sigma}
\end{eqnarray}

\noindent The $T^{2}(1-T)$ dependence results in minima in the amplitude of the 
voltage dependent fluctuations in the conductance at $T=0$ or $T=1$ and a 
maximum at $T=2/3$. This result can be extended to multiple conducting modes, 
when we assume that the probability to be scattered back to the contact is 
independent of the mode index, i.e. that defects scatter a wave equally into all 
available modes. The term $T^2(1-T)$ in Eq.\,(\ref{sigma}) is replaced for the 
$N$-mode problem by $\sum^{N}_{n=1}T_{n}^{2}(1-T_{n})$. 
\begin{figure}
\begin{center}
\leavevmode
\epsfig{figure=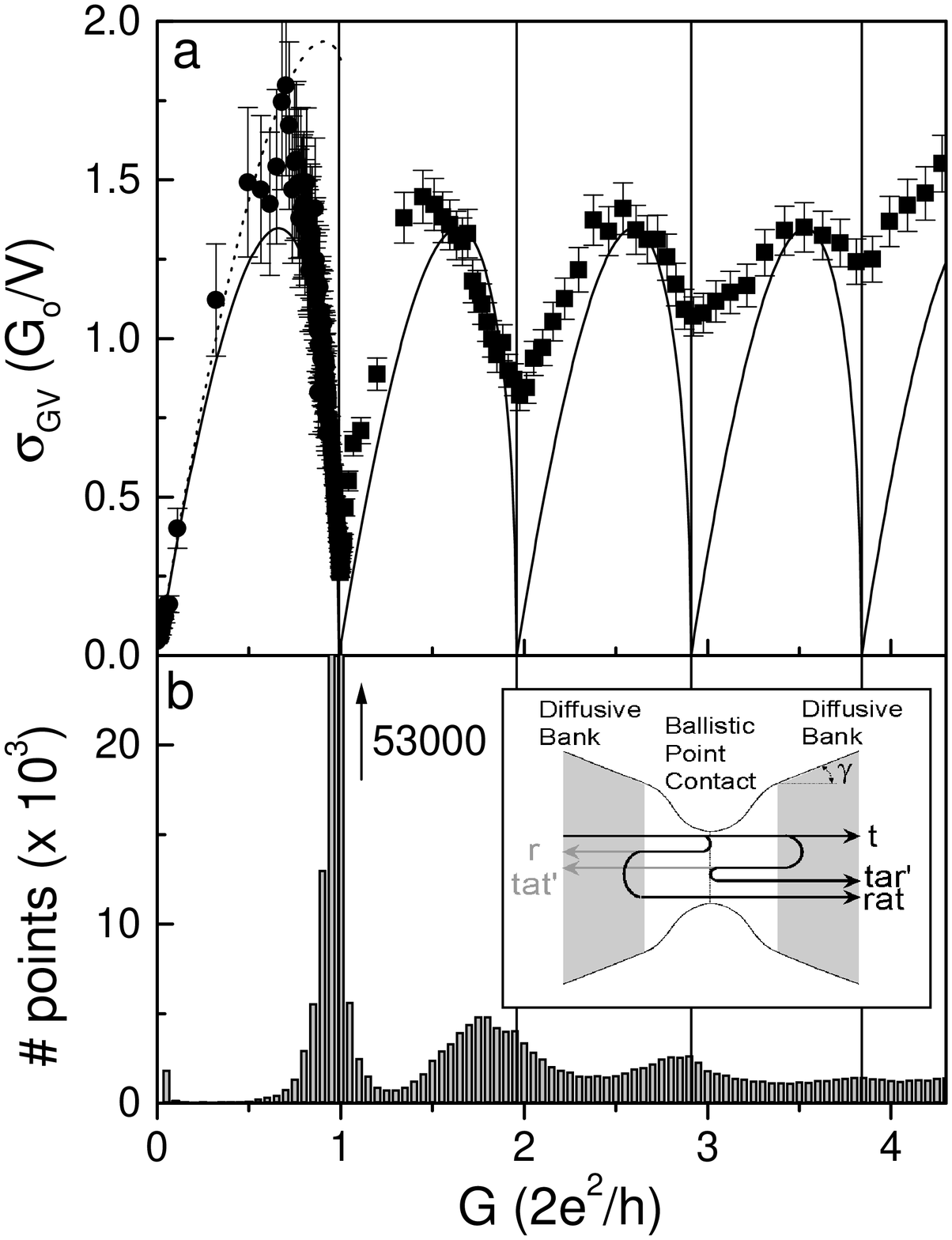,width=12cm,angle=0}
\end{center}
\caption{(a) Standard deviation of the voltage dependence of the conductance 
versus conductance for 3500 curves. The circles are the averages for 300 points, 
and the squares for 2500 points. The solid and dashed curves depict the 
calculated behavior for a single partially open channel and a random 
distribution over two channels, respectively. The vertical gray lines are the 
corrected integer conductance values (see text). (b) Conductance histogram 
obtained from the same data set. The peak in the conductance histogram at 
$G_{0}$ extends to 53000 on the y-scale. (Inset) Schematic diagram of the 
configuration used in the analysis. The dark lines with arrows show the paths, 
which contribute to the conductance fluctuations in lowest order.} 
\label{fig:fig3} \end{figure}
When comparing the experimental data for $\sigma_{GV}$ with our theoretical 
model, we need to be aware that the experimental data have been sorted according 
to their conductance value. A given value for $G=G_0\sum T_n$ can be constructed 
in many ways from a choice of transmission values $\{ T_n\}$. The experimental 
values for $\sigma_{GV}$ are, therefore, an average over impurity configurations 
{\it and} transmission values. Assuming these averages are independent, we can 
compare the data with various choices for the distribution of the transmissions. 
The dashed curve in Fig.\,3a shows the behavior of $\sigma_{GV}$ for a random 
distribution of two $T_n$'s in the interval $\{0,1\}$ under the constraint $T_1 
+ T_2 = G/G_0$, where the amplitude has been adjusted to fit the data. 
Alternatively, the full curve shows the behavior for a single partially open 
channel, i.e. in the interval $G/G_0=\{0,1\}$ there is a single channel, in 
$\{1,2\}$ there are two channels with one fully open, etc. The latter 
description works surprisingly well, in particular for the minimum near 1\, 
$G_0$, and for the fact that the maxima are all nearly equal. 

Note that the minima in Fig.\,3a are found slightly below the integer values. A 
reduction of the conductance with respect to the bare conductance of the 
contact, $G_0\sum T_n$, results from total probability for back scattering on 
the same defects which give rise to the fluctuations. We can estimate the 
correction as the sum over incoming channels, $n$, and their probability to 
return via any channel, $m$,  $\sum T_n - G/G_0 = 2 \sum_{n,m}  T_n T_m \langle 
|a_{nm}(E=0)|^2\rangle$. The total return probability $\langle 
|a_{nm}|^2\rangle$ we approximate by the substitution $\int_{\tau_e}^{\infty} 
P_{cl}(\tau) \text{d}\tau $. Thus we expect a correction term $G = G_{0}(\sum 
T_{n} - 2\langle |a_{nm}|^2\rangle (\sum T_{n})^{2}$). In Fig.\,3 the vertical 
gray lines indicate the shift below integer values for $\langle 
|a_{nm}|^2\rangle = 0.005$, which is equivalent to a classical series resistance 
of 130\,$\Omega$. From this value for $\langle |a_{nm}|^2\rangle$ we obtain an 
estimate for $l_e = 5$\,nm, which is consistent with the value 
obtained from the fluctuation amplitudes discussed below.

In our experiment we measure the second derivative with a modulation amplitude 
of 20\,mV. This limits the path lengths to which we are sensitive to those 
smaller than  $\sim$100\,nm ($L_V=v_{F} \hbar / eV$). From the amplitude of the 
full curve in Fig.\,3a we obtain an estimate of $l_{e} = 5\pm 1$\,nm, assuming 
reasonable values for the opening angle $\gamma$ of 30$^{o}$-50$^{o}$ 
\cite{agrait2}. This value is consistent with our assumption that $d \ll l_e \ll 
L_V$, where $d$ is the contact diameter. The estimate for $l_e$ is sensitive to 
the functional form of the factors in front of the $T^2(1-T)$-term in Eq.\,(2), 
which was not tested in detail. Measurements of the dependence on modulation 
amplitude $V$ are under way. However, the thermopower of atomic size gold 
contacts was recently measured \cite{me} and has been found to be determined by 
the same mechanism, but it was measured on an energy scale nearly two orders of 
magnitude lower. It gives the same estimate of $l_e=5\pm 1$\,nm, consistent 
with the present value. 

Conductance fluctuations \cite{spivak} have been observed previously in ballistic contacts 
with diameters an order of magnitude larger compared to our contacts, and were 
measured as a function of both magnetic field and bias voltage \cite{holweg}. In 
that work, the quantum suppression of the fluctuations which we report here, is 
not observable due to the fact that many nearly-open channels contribute to the 
conductance for large contacts. The model introduced by Kozub {\it et al.} 
\cite{kozub} to describe the results of Ref.\,\onlinecite{holweg} contains only 
terms due to the interference of two diffusing trajectories, which are second 
order in $|a|^2$. 

The minimum observed at $G_{0}$ in Fig.\,3a is very sharp, close to the full 
suppression of fluctuations predicted for the case of a single channel. To 
describe the small deviation from zero, it is sufficient to assume that there is 
a second channel which is weakly transmitted, $T_2\ll 1$, and $T_1 \simeq 1$ 
such that $T_1 + T_2 = 1$. For this case it is easy to show that the value of 
$\sigma_{GV}^2$ at the minimum is proportional to the average value of $T_2$. We 
obtain $\langle T_2\rangle=0.005$, implying that, on average only 0.5\,\% of the 
current is carried by the second channel. For the minima near 2, 3 and 4\,$G_0$ 
we obtain higher values: 6, 10 and 15\,\%, respectively. 

The well-developed structure observed in $\sigma_{GV}$, with a dependence which 
closely follows the $\sqrt{\sum T_n^2(1-T_n)}$ behavior of Eq.\,2, demonstrates 
a property of the contacts which we refer to as the saturation of transmission 
channels: There is a strong tendency for the channels contributing to the 
conductance of atomic-size gold contacts to be fully transmitting, with the 
exception of one, which then carries the remaining fractional conductance. 
Fig.\,3 shows that the positions of the minima in $\sigma_{GV}$ do not all 
coincide with those of the maxima in the histogram. This is most pronounced for 
the feature below $G=2$\,$G_0$. We conclude that the statistically preferred 
values of the histograms do not necessarily correspond with perfect transmission 
of the bare contact. We propose that the appearance of peaks in the histograms, 
such as the one at 1.75\,$G_0$, arises from preferred atomic configurations. 

The concept of the saturation of transmission channels is consistent with recent 
work, which shows that, for monovalent metals, the conductance at $G=1$\,$G_0$ 
of a single atom is carried by a single mode \cite{cuevas,subgap,vdBrom}. Conversely, 
based on the analysis of the subgap structure for superconducting aluminum by 
Scheer {\it et al.} \cite{subgap,scheer}, which showed that typically three 
channels contribute to the conductance at $G=1$\,$G_0$, we should expect that 
aluminum does not show a pronounced suppression of conductance fluctuations near 
integer values. Indeed, preliminary measurements of $\sigma_{GV}$ on this $p$-
metal, exhibit results for $G\leq G_0$ that are close to a random distribution 
over 3 transmission channels, while the monovalent metals Ag and Cu show 
behavior similar to Au. 

This work is part of the research program of the ''\nobreak{Stichting} FOM'', 
which is financially supported by NWO. BL and JMvR acknowledge the stimulating 
support of L.J. de Jongh, and we thank E.Scheer and J.Caro for helpful 
discussions.



\begin{thebibliography}{99}
\bibitem{daniel} Strictly, we should consider the probability until the first 
return to the contact, but we estimate the correction to be of order 
$(d/l_{e})$, with $d$ the contact diameter.
\end{thebibliography}
\end{document}